# Minimizing Human-Induced Variability in Quantitative Angiography for Robust and Explainable AI-Based Occlusion Prediction

Parmita Mondal[1,2], Mohammad Mahdi Shiraz Bhurwani[3], Swetadri Vasan Setlur Nagesh[2,4], Pui Man Rosalind Lai[1,2], Jason Davies[1,2], Elad Levy[1,2], Kunal Vakharia[5], Michael R Levitt[6], Adnan H Siddiqui[1,2,4], Ciprian N Ionita[1,2,3]

[1]Department of Biomedical Engineering, University at Buffalo, Buffalo, NY 14260
[2]Canon Stroke and Vascular Research Center, Buffalo, NY 14203
[3]QAS.AI Inc, Buffalo, NY 14203
[4]University at Buffalo Neurosurgery, Inc, Buffalo, NY 14203
[5]Department of Neurosurgery and Brain Repair, University of South Florida, Tampa, FL 33609
[6]Department of Neurological Surgery, University of Washington, Seattle, WA 98195
Correspondence to Dr Ciprian N Ionita, Biomedical Engineering, University at Buffalo, State University of New York, Buffalo, NY 14203, USA; cnionita@buffalo.edu

Correspondence:  Ciprian N. Ionita, Department of Biomedical Engineering, University at Buffalo, email: cnionita@buffalo.edu

**ABSTRACT**

**Background:** Bias from contrast injection variability is a significant obstacle to accurate intracranial aneurysm (IA) occlusion prediction using quantitative angiography (QA) and deep neural networks (DNNs). This study explores bias removal and explainable AI (XAI) for outcome prediction.

**Objective**: Implement injection bias removal algorithm for reducing QA variability and examine XAI's impact on the reliability and interpretability of deep learning models for occlusion prediction in flow diverter-treated aneurysms.

**Materials and Methods:** This study used angiograms from 458 patients with flow diverters-treated IAs with six-month follow-up defining occlusion status.  We minimized injection variability by deconvolving the parent artery input to isolate the aneurysm's impulse response, then reconvolving it with a standardized injection curve. A deep neural network (DNN) trained on these QA-derived biomarkers predicted six-month occlusion. Local Interpretable Model-Agnostic Explanations (LIME) identified the key imaging features influencing the model, ensuring transparency and clinical relevance.

**Results:** The DNN trained with uncorrected QA parameters achieved a mean area under the receiver operating characteristic curve (AUROC) of 0.60±0.05 and an accuracy of 0.58±0.03. After correcting for injection bias by deconvolving the parent artery input and reconvolving it with a standardized injection curve, the DNN's AUCROC increased to 0.79±0.02 and accuracy to 0.73±0.01. Sensitivity and specificity were 67.61±1.93% and 76.19±1.12%, respectively. LIME plots were added for each prediction to enhance interpretability.

**Conclusions:** Standardizing QA parameters via injection bias correction improves occlusion prediction accuracy for flow diverter-treated IAs. Adding explainable AI (e.g., LIME) clarifies model decisions, demonstrating the feasibility of clinically interpretable AI-based outcome prediction.

 **Keywords:** Occlusion, Intracranial Aneurysm, Injection Bias Correction, Quantitative Angiography

# INTRODUCTION

Flow diverters have become a widely used device in treating intracranial aneurysms (IAs), offering an alternative to coil embolization or microsurgery.(1) However, despite their effectiveness, complete IA healing is not always achieved, leading to variable treatment success. Approximately 20-30% of patients show incomplete occlusion at six months, requiring extended follow-up and potential retreatment.(2-6) The need for retreatment increases procedural risks and imposes significant healthcare costs, particularly in cases requiring additional interventions. During this period, patients remain at risk, as residual aneurysms have been linked to increased likelihood of delayed hemorrhage, with risk levels varying on IA size, location, and treatment approach.(7, 8) The need for retreatment not only elevates procedural risks but also adds substantial healthcare costs, especially in cases requiring additional interventions. Reliable occlusion prognosis could optimize patient management, identify cases needing early intervention, and reduce unnecessary follow-ups, limit health expenditures, and ultimately improving clinical outcomes.

During IA treatment, neuro-interventionalists typically rely on a qualitative evaluation of treatment success based on the intra-arterial flow of the contrast agent via digital subtraction angiography (DSA). (9, 10) Although DSA has been traditionally used for structural assessment, more recently two-dimensional quantitative angiography (2D-QA) has been studied to quantify hemodynamic conditions associated with endovascular outcomes.(11-13) 2D-QA involves analyzing contrast behavior by synthesizing time density curves (TDCs) from specified regions of interest (ROIs) and extracting imaging biomarkers from these curves.(14, 15)

AI-based models leveraging 2D-QA parameters have shown promise in occlusion prediction after IA treatment. Deep neural networks (DNNs) trained on QA-derived imaging biomarkers have demonstrated strong predictive performance.(16, 17) Differences in injection techniques during DSA (in particular, hand-injection of contrast) introduce inconsistencies in QA parameters, affecting AI model reliability.(18) Prior work using normalization techniques has mitigated these issues with reasonable success,(14, 15) but doesn't permit comparisons across patients, neuro-interventionalist or institutions, reducing clinician confidence and limiting 2D-QA application.

In this study, we propose singular value decomposition (SVD)-based deconvolution to standardize QA parameters and reduce hand-injection variability, thereby enhancing the reliability of AI-based IA occlusion prediction. By ensuring consistent QA reports across patients, we expect to improve model performance and facilitate the use of model interpretability techniques such as Local Interpretable-Agnostic Explanations (LIME).(19)

# MATERIAL AND METHODS

## Data collection

Data collection and analysis were approved by our Institutional Review Board (IRB). The inclusion schema for the study is shown in Figure 1. Each patient was treated with a flow diverter, and each patient underwent follow-up imaging at least six months after the date of treatment.

We recorded DSA sequences for each case at three different time points: pre-treatment, post-treatment, and follow-up. Occlusion outcome (occluded/unoccluded) was provided by neurosurgeons from the follow-up DSAs, based on their clinical judgement. For each case, neurosurgeon fellows and residents (having 2 years of neuro-intervention experience) drew regions of interest (ROI) over the aneurysm and their corresponding main artery (inlet) to generate time TDCs, from which QA parameters were computed.

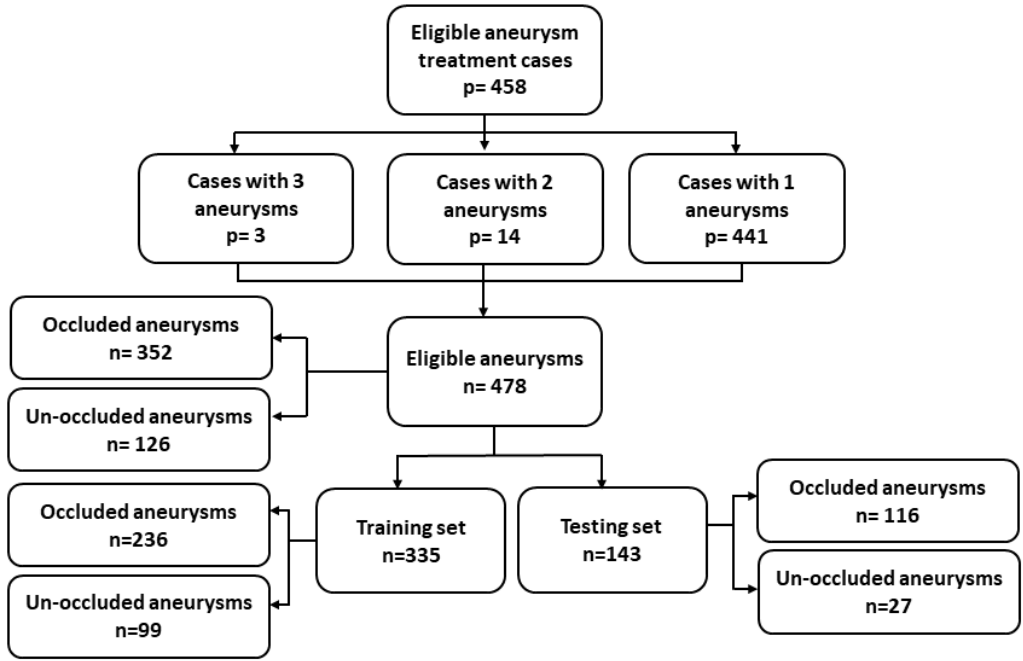

Figure 1. Study inclusion schema: p refers to the number of patient and n refers to the number of aneurysms.

**Quantitative angiographic parameter analysis using deconvolution methods**

To minimize variability in QA parameters caused by differences in contrast injection techniques, we applied a mathematical approach known as singular value decomposition (SVD) to each DSA.(20) Specifically, we analyzed the contrast concentration measured at the vessel inlet, known as the arterial input function (AIF), to isolate the aneurysm's intrinsic flow characteristics from the injection technique itself.(21) In this method, the AIF is structured into a specialized matrix format (Toeplitz matrix), which allows separation (deconvolution) of the aneurysm's specific response to the injected contrast. To reduce the influence of noise in this calculation, smaller, less meaningful components (singular values below a certain threshold) were removed. The outcome of this process is the aneurysm's impulse response function (IRF), representing how the aneurysm inherently responds to a standardized injection of contrast (Figure 2). Finally, we re-convolved the IRF with a standardized injection profile to obtain a clinically relevant and injection-invariant TDC ($Q_{new}$) of the IA dome. This final step renders all angiograms comparable as if the same injector had been used (Figure 2).

From $Q_{new}$, we extracted QA parameters, including Bolus Arrival Time (BAT), Mean Transit Time (MTT), Time to Peak (TTP), Peak Height (PH), Area Under the Curve (AUC), and Maximum Derivative (Max-Df)(22). BAT is defined as the first point where contrast intensity exceeds 1% of its peak value. TTP corresponds to the time at which the contrast reaches its maximum intensity (PH). MTT is computed using the full width at half maximum (FWHM), with half-maximum defined as PH/2. Max-Df represents the peak derivative between BAT and TTP, quantifying the rate of contrast inflow. AUC is obtained by integrating the TDC over the entire acquisition period, and additional AUC values (AUC-0.5, AUC-1.0, AUC-1.5, AUC-2.0) are computed over time intervals defined as 0.5×, 1×, 1.5×, and 2× the MTT to capture temporal contrast flow dynamics. Additionally, we computed the cross-correlation index (Cor) between the standardized AIF and the injection-invariant Qnew within the aneurysm dome, providing a measure of how closely the aneurysm's hemodynamic response follows the inlet contrast dynamic. Finally, the parameters were reported as a ratio of the post-device $Q_{new}$ parameters to the baseline corresponding parameters.

## Occlusion outcome prediction using deep neural network

We developed a deep neural network (DNN) using Keras to predict IA occlusion as a six-month binary outcome (occluded vs. unoccluded).(23) The model was trained on a dataset containing QA-new parameters extracted from dynamic DSA sequences. The dataset was stratified and split into training (80%) and testing (20%) subsets, followed by an additional division of the training set into training (80%) and validation (20%) cohorts.

The DNN architecture consisted of an input layer with a number of neurons corresponding to the selected features, followed by one hidden layer. The input and the hidden layers each contained 256 neurons with Swish activation, and L1-L2 regularization to mitigate overfitting.(24) A dropout layer with a 10% rate was applied after the hidden layer. The output layer contained a single neuron with sigmoid activation for binary classification. The model was compiled using the Adam optimizer with a learning rate of 0.001 and binary cross-entropy loss.(25) Training was conducted for a maximum of 300 epochs with early stopping based on validation loss (patience = 15). Additionally, a learning rate reduction strategy was applied, decreasing the learning rate by a factor of 0.1 after three consecutive epochs without improvement.

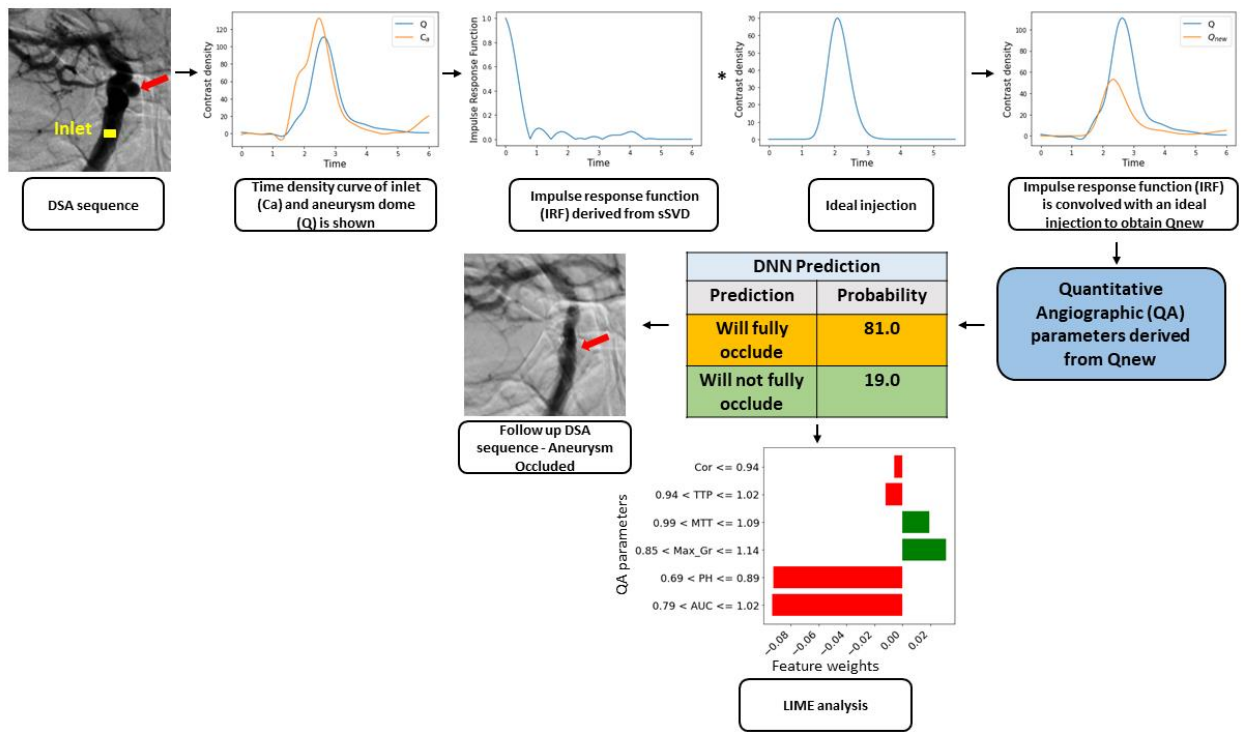

**Figure 2.** Study workflow. Top row, left to right: The DSA sequence (left) is used to extract the time density curve (TDC) of the inlet ($C_a$; yellow square) and TDC of the IA dome (Q; red arrow). Next the impulse response function (IRF) is derived from single value decomposition (SVD) convolved with an ideal injection, to generate ($Q_{new}$). Finally, the quantitative angiographic (QA) parameters derived from $Q_{new}$ and deep neural network (DNN) are used to predict IA occlusion. LIME analysis is shown at the bottom. To avoid data overcrowding in the LIME analysis, only the top six parameters with the largest contributions to the model's prediction were retained.

To evaluate model robustness, Monte Carlo Cross Validation (MCCV) was performed with 10 random train-test splits.(26) In each iteration, the model was trained and tested on a new split, and performance was assessed using classification accuracy, receiver operating characteristic (ROC) curve analysis, area under the ROC curve (AUROC), and confusion matrices. Mean and standard deviations for all metrics were calculated across the 10-fold MCCV.

To enhance interpretability, Local Interpretable Model-Agnostic Explanations (LIME) was applied to assess feature contributions for individual predictions. LIME is a technique used to explain the predictions of DNN models by approximating complex models with interpretable local surrogate models,(27) (Figure 2). The LIME analysis at the bottom of the figure provides an interpretation of how individual QA parameters influenced the DNN outcome. LIME works by locally perturbing the trained model, meaning that small variations are introduced to the case-specific parameters to observe how they affect the model's prognosis. The DNN in this case predicts the probability of the undesired event—failure to occlude—where a value of 1 indicates failure to occlude and a value of 0 indicates successful occlusion. The LIME plot starts with an initial probability based on chance, then evaluates how each parameter shifts the prediction toward either outcome. The x-axis represents the contribution of each parameter to the final prediction, with positive values increasing the probability of failure to occlude and negative values decreasing it.

In Figure 2 specific case, we observe that lower PH and AUC relative values, represented by post-/pre-device placement ratios of less than one, shift the prediction toward successful occlusion. This suggests that less contrast is entering the aneurysm dome, likely indicating a reduction in inflow. Conversely, the lower MTT value pulls the prediction toward occlusion failure, suggesting that contrast is spending less time in the aneurysm dome. This could imply faster washout due to increased outflow at the aneurysm neck, even though the overall contrast entry into the dome is reduced.

## Statistical analysis

For clarity, we will refer to the data without injection bias removal as 'no-SVD' and the data with injection bias removal as 'with SVD.' Statistical analysis was performed on each QA parameter (Cor, TTP, MTT, PH, AUC, Max-Gr), categorized into their respective sub-groups: relative no-SVD (ratio of post-treatment to pre-treatment QA parameters no-SVD) and relative with SVD (ratio of post-treatment QA parameters to pre-treatment QA parameters with SVD). These parameters were analyzed in relation to treatment status and IA occlusion outcome (occluded vs. unoccluded). Box and whisker plots were generated for each parameter within its respective subgroup to illustrate differences between occluded and unoccluded IA cases. Additionally, the AUROC was computed for each QA parameter to assess its predictive value for IA occlusion outcome.

## RESULTS

We gathered data from 458 patients with 478 IAs who underwent flow diverter treatment at a single institution. The extracted QA parameters for each IA were analyzed in relation to occlusion outcomes, with distributions presented as box-and-whisker plots in Figure 3. These plots illustrate the variability of parameters between occluded and unoccluded aneurysms, both with and without SVD-based correction.

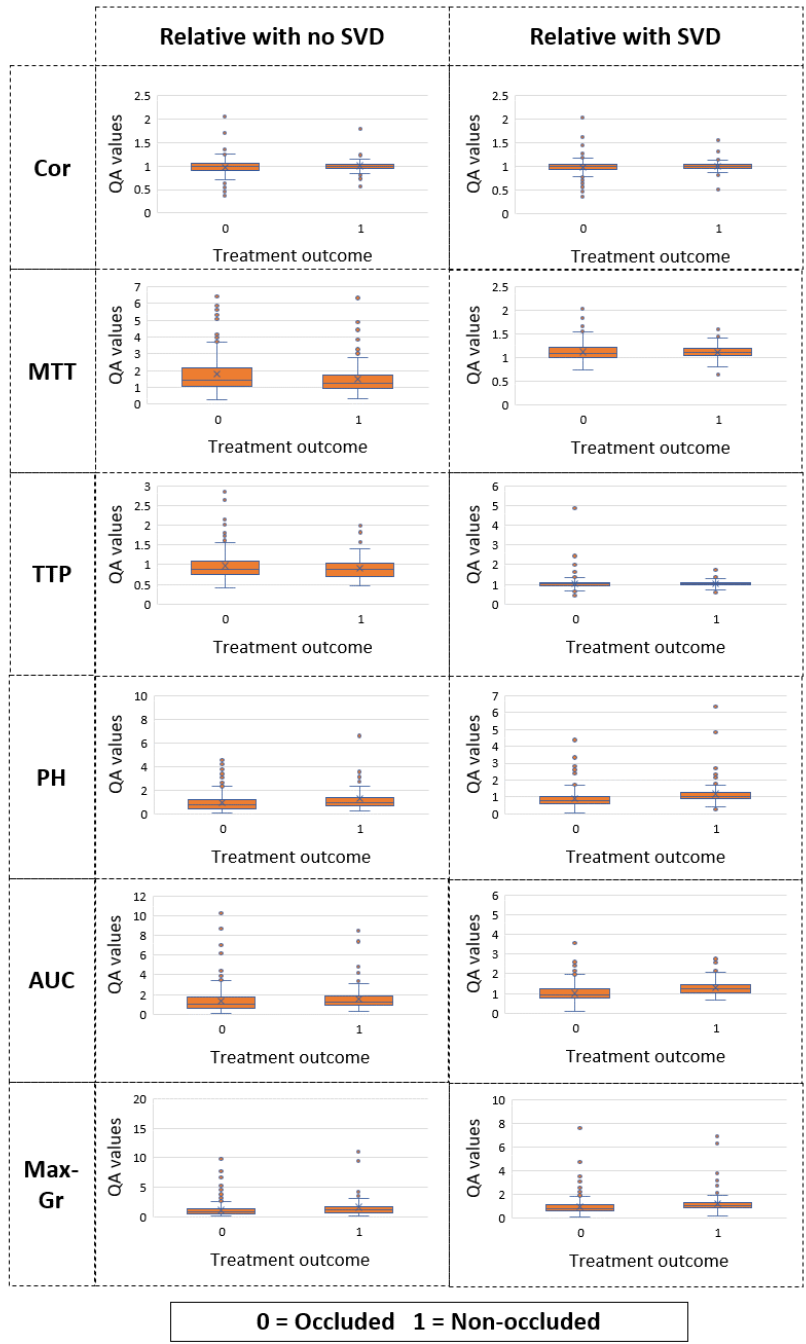

**Figure 3.** Box and whisker plot for each QA parameter for relative data (ratio of post-treatment and pre-treatment QA parameters) with no SVD and SVD, is shown.

The uncorrected parameters exhibited wide variability, whereas SVD-standardized parameters demonstrated a more compact distribution, suggesting that the deconvolution approach effectively reduced injection-related inconsistencies.

Statistical comparisons between occluded and unoccluded cases indicated that individual QA parameters alone had limited predictive value, with overlapping distributions in several cases. However, relative QA parameters—computed as the ratio of post-treatment to pre-treatment values—showed stronger differentiation between the two groups. These findings highlight the importance of normalizing injection effects before using QA metrics for occlusion prediction.

The DNN trained with no-SVD-corrected QA parameters achieved a mean AUROC of $0.60 \pm 0.05$ and a classification accuracy of $0.58 \pm 0.03$. The DNN trained on SVD-corrected QA parameters achieved a mean AUROC of $0.79 \pm 0.02$ and a classification accuracy of $0.73 \pm 0.01$ (Table 1). Sensitivity and specificity were $67.61 \pm 1.93\%$ and $76.19 \pm 1.12\%$, respectively, indicating that the model distinguished occluded from unoccluded IAs with reasonable reliability. Comparisons across different feature sets revealed that models trained on SVD-corrected parameters outperformed those using raw QA parameters, confirming the benefits of standardization.

The AUROC values of individual QA parameters were lower than that of the DNNs incorporating multiple features. Specifically, PH and AUC emerged as the strongest individual predictors, whereas TTP, MTT, and Max-Df contributed less to the model's decision-making. However, none of the parameters alone reached the predictive performance of the combined feature set.

**Table 1:** Performance of DNN in predicting occlusion outcome of intracranial aneurysms in the form of average classification accuracies and average AUROCs along with their standard deviation (SD).

| QA parameters | Sub-group | Accuracy ± SD (%) | AUROC ± SD (%) |
|---|---|---|---|
| Only Cor | Pre-treatment with no SVD | N/A | 0.56 ± 0.03 |
| | Pre-treatment with SVD | | 0.49 ± 0.03 |
| | Post-treatment with no SVD | | 0.57 ± 0.03 |
| | Post-treatment with SVD | | 0.58 ± 0.03 |
| | Relative | | 0.58± 0.03 |
| Only TTP | Pre-treatment with no SVD | | 0.46 ± 0.03 |
| | Pre-treatment with SVD | | 0.48 ± 0.03 |
| | Post-treatment with no SVD | | 0.46 ± 0.03 |
| | Post-treatment with SVD | | 0.50 ± 0.02 |
| | Relative | | 0.50 ± 0.03 |
| Only MTT | Pre-treatment with no SVD | | 0.44 ± 0.03 |
| | Pre-treatment with SVD | | 0.48 ± 0.03 |
| | Post-treatment with no SVD | | 0.41 ± 0.03 |
| | Post-treatment with SVD | | 0.45 ± 0.02 |
| | Relative | | 0.51 ± 0.03 |
| Only PH | Pre-treatment with no SVD | | 0.49 ± 0.03 |
| | Pre-treatment with SVD | | 0.54 ± 0.02 |
| | Post-treatment with no SVD | | 0.60 ± 0.02 |
| | Post-treatment with SVD | | 0.63 ± 0.02 |
| | Relative | | *0.71* ± 0.02 |
| Only AUC | Pre-treatment with no SVD | | 0.44 ± 0.03 |
| | Pre-treatment with SVD | | 0.52 ± 0.02 |
| | Post-treatment with no SVD | | 0.53 ± 0.02 |
| | Post-treatment with SVD | | 0.60 ± 0.02 |
| | Relative | | 0.73 ± 0.02 |
| Only Max-Gr | Pre-treatment with no SVD | | 0.49 ± 0.03 |
| | Pre-treatment with SVD | | 0.55 ± 0.02 |
| | Post-treatment with no SVD | | 0.61 ± 0.02 |
| | Post-treatment with SVD | | 0.63 ± 0.02 |
| | Relative | | 0.68 ± 0.02 |
| Cor, TTP, MTT, PH, AUC, Max-Gr | Pre-treatment with no SVD | 0.55 ± 0.02 | 0.51 ± 0.05 |
| | Pre-treatment with SVD | 0.52 ± 0.02 | 0.56 ± 0.02 |
| | Post-treatment with no SVD | 0.57 ± 0.02 | 0.59 ± 0.04 |
| | Post-treatment with SVD | 0.62 ± 0.02 | 0.65 ± 0.03 |
| | Relative | 0.71 ± 0.01 | 0.79 ± 0.02 |
| Cor, TTP, MTT, PH, AUC, Max-Gr, AUC-0.5, AUC-1.0, AUC-1.5, AUC-2.0 | Pre-treatment with no SVD | 0.51 ± 0.02 | 0.50 ± 0.05 |
| | Pre-treatment with SVD | 0.54 ± 0.03 | 0.57 ± 0.03 |
| | Post-treatment with no SVD | 0.58 ± 0.02 | 0.58 ± 0.04 |
| | Post-treatment with SVD | 0.61 ± 0.02 | 0.64 ± 0.03 |
| | Relative | 0.73 ± 0.01 | 0.79 ± 0.02 |

To assess model interpretability, LIME analysis was applied to every patient in the testing cohort, providing patient-specific insights into the decision-making process of DNN. The LIME results (Figure 2) revealed that AUC and PH positively influenced occlusion predictions, while TTP and Max-Df had negative contributions. Notably, the relative importance of each parameter varied from patient to patient, indicating that the model's predictions were individualized rather than relying on a single dominant factor. This suggests that higher AUC and PH values are generally associated with occlusion, whereas delayed contrast transit (higher TTP) may indicate persistent flow within the aneurysm.

**DISCUSSIONS**

In this study, we reinforced and improved on previous findings that QA-derived imaging biomarkers can predict aneurysm occlusion using a DNN model.(14, 28, 29) To address a key limitation – variability in contrast injection technique – we incorporated SVD-based deconvolution. Our results demonstrate that SVD correction improves the reliability of QA parameters, leading to enhanced model performance and a more standardized approach to occlusion prediction. Furthermore, by integrating explainable AI (XAI) techniques such as LIME, we provide a framework for individualized, interpretable predictions, making AI-driven decision support more applicable to clinical practice.

One of the contributions of this study is the validation of SVD-based correction as a preprocessing step to mitigate injection variability. Prior work has shown that QA biomarkers derived from DSA are predictive of occlusion, but inconsistencies in injection technique introduce variability that affects model performance. We demonstrated that standardizing contrast injection profiles using SVD deconvolution leads to more compact distributions of QA parameters, as shown in Figure 3, reducing noise and improving predictive accuracy. The DNN trained on SVD-corrected data achieved an AUROC of $0.79 \pm 0.02$ and a classification accuracy of $0.73 \pm 0.01$, outperforming models trained on uncorrected QA parameters. Moreover, occlusion prediction benefits from combining multiple QA parameters rather than relying on a single biomarker. While AUC and PH emerged as the strongest individual predictors, their standalone performance was inferior to the full multivariable model. Conversely, parameters such as TTP and Max-Df negatively contributed to occlusion predictions, suggesting that delayed contrast transit and lower peak gradients may be indicative of persistent IA perfusion. These findings reinforce the need for multivariate modeling to capture the complex hemodynamic interactions that influence occlusion outcomes.

To enhance model interpretability, we incorporated LIME analysis to extract case-specific insights into the DNN's decision-making process. The contribution of each QA parameter varied across patients, confirming that the model does not operate on a single dominant feature but dynamically adjusts weight distributions based on individual patient data. This patient-specific approach to explainability is critical for clinical adoption, as it allows neuro-interventionalists to assess why the model predicts occlusion or persistence in a given case, rather than accepting the output as a "black box" decision. Additionally, LIME facilitates the rejection of model predictions that contradict established hemodynamic principles. If the model generates a biologically implausible prediction, it can be flagged and discarded, thereby improving the reliability of AI-assisted decision-making.

Interpretability in LIME relies on the consistency of data input, reinforcing the need for injection bias correction via SVD. Without this correction, variations in contrast injection introduce uncontrolled drift in QA parameters, leading to interpretative inconsistencies. SVD correction ensures that LIME explanations reflect physiological changes rather than injection-related artifacts, making the results clinically meaningful. However, preserving interpretability requires consistency beyond training. If inference data is processed differently than training data, discrepancies could emerge, leading again to misleading interpretations. In summary, while SVD accounts for injection variability, preprocessing steps inconsistent with those used in the model training, must be avoided to ensure the reliability of LIME-based explanations.

The clinical impact of this framework extends beyond prognostication, offering real-time decision support during interventions. Neuro-interventionalists could use the reported values from the ideal QA to assess the contributing factors behind the model's prediction and decide whether to change the treatment strategy. For instance, if the model predicts a low likelihood of occlusion and attributes it to short MTT or increased PH, it may indicate persistent/strong inflow within the IA. In such cases, neuro-interventionalists can evaluate whether the prediction aligns with their assessment of flow-diverter positioning, ensuring there are no kinks or wall apposition issues affecting flow modification. If wall apposition is inadequate, balloon angioplasty can be performed to optimize contact between the device and the vessel wall. If the flow disruption remains insufficient, an additional flow diverter may be deployed to enhance metal coverage and improve treatment efficacy. Because LIME-derived feature importance values vary by patient, this AI-driven decision-support system provides individualized insights, enabling clinicians to interpret the factors driving the model's decision and determine whether intraoperative adjustments are necessary.

Beyond intraoperative use, this methodology could be extrapolated to pre-procedural planning, particularly in device sizing. By simulating different device configurations and predicting their impact on occlusion, neuro-interventionalists could make more informed choices regarding flow diverter selection before deployment. This could reduce procedural variability and improve long-term outcomes by ensuring that the chosen device has the optimal metal coverage and hemodynamic effect.

One of the limitations of this study is foreshortening due to the 2D nature of the imaging system, since cerebral arteries are complex 3D structures represented in DSA by 2D planar images(30). One way to address this limitation is to utilize 3D QA methods which could eliminate many foreshortening errors. Additionally, patient demographic and clinical characteristics that may influence treatment outcomes were not incorporated into our model. Finally, our study was limited by the relatively small sample size, single-center retrospective data collection and self-adjudicated outcomes. Further large-scale study across multiple institutions would enhance the robustness of the DNN, likely improving accuracy to a clinically-acceptable level.

## CONCLUSIONS

We developed an SVD-based deconvolution method to standardize QA parameters, mitigating variability introduced by manual contrast injection. By integrating this preprocessing step with a deep neural network, we enhanced the accuracy and reliability of IA occlusion prediction following flow-diverter treatment. This approach represents a novel strategy for harmonizing angiographic data, enabling more consistent prognostication. Additionally, by incorporating explainable AI techniques, we introduced a clinically interpretable decision-support system, equipping neuro-interventionalists with real-time insights to assess model-driven predictions and make informed intraoperative adjustments. Future work will focus on multi-center validation and integration with multimodal imaging, further advancing AI-driven decision-making in cerebrovascular interventions.

## ACKNOWLEDGEMENTS

This work was supported by NSF SBIR Award #2304388

## COMPETING INTERESTS

**Parmita Mondal**
No conflict of interest to disclose

**Mohammad Mahdi Shiraz Bhurwani, PhD**
Employee and Shareholder at QAS.AI Inc, PI on NSF SBIR Award #2304388.

**Swetadri Vasan Setlur Nagesh**
UB award from QAS.AI, Parent NSF Phase 2 STTR # 2111865, Grant from University at Buffalo Center for Advanced Technology in Big Data and Health Sciences (UB-CAT)

**Pui Man Rosalind Lai**
No conflict of interest to disclose.

**Jason Davies**
Research Grants: NIH R21/R01, NSF SBIR, UB-CAT, Buffalo Translational Consortium, Cummings Foundation, NVidia, Google

Financial Interest: QAS.ai, Rist Neurovascular, Cerebrotech, Synchron, Hyperion, Kandu, Radical Catheter Technologies, ASCEND

Consultant/Advisory Board: Medtronic, Microvention, Imperative Care, Xenter, RapidPulse, Rapid Medical, Canon, J&J

National PI/Steering Committees: StrokeNET DSMB, EMBOLISE, SUCCESS, RapidPulse, SBIR/STTR, NIH NINDS/NLM Study sections

**Elad Levy**

Consulting fees: Clarion, GLG Consulting, Guidepoint Global, Medtronic, StimMed, Mosaic; Payment or honoraria for lectures, presentations, speakers' bureaus, manuscript writing or educational events: Medtronic, Penumbra, MicroVention (now Terumo Neuro), Integra; Patents planned, issued, or pending: Ultrasonic Surgical Blade;

Participation on a Data Safety Monitoring Board or Advisory Board: NeXtGen Biologics, Cognition Medical; Endostream Medical, IRRAS AB; Leadership or fiduciary role in other board, society, committee or advocacy group, paid or unpaid: CNS, ABNS, UBNS;

Stock or stock options (shareholder or ownership interest): NeXtGen Biologics, RAPID Medical, Claret Medical, Cognition Medical, Imperative Care, StimMed, Three Rivers Medical, Q'Apel, Dendrite;

Other financial or non-financial interests: Haniva Medical Technology (Chief Medical Officer); Medtronic (National PI: Steering Committees for SWIFT Prime and SWIFT Direct trials; SHIELD trial; Site PI: STRATIS Study – Sub I); Penumbra (National PI: THUNDER trial); MicroVention (now Terumo Neuro) (Site PI: CONFIDENCE Study).

**Kunal Vakharia**
University Grants/Research Support from NSF SBIR grant; Consultant Fees from Medtronic, Terumo Neuro, Stryker, Phenox Wallaby; Stock/Shareholder in Von Vascular

**Michael R Levitt**
Unrestricted educational grants from Medtronic and Stryker; consulting agreement with Aeaean Advisers, Metis Innovative, Genomadix, AIDoc and Arsenal Medical; equity interest in Proprio, Stroke Diagnostics, Apertur, Stereotaxis, Fluid Biomed, Synchron and Hyperion Surgical; editorial board of Journal of NeuroInterventional Surgery; Data safety monitoring board of Arsenal Medical.

**Adnan H. Siddiqui MD, PhD**
Current Research Grants: Co-investigator for NIH - 1R01EB030092-01, Project Title: High Speed Angiography at 1000 frames per second; Mentor for Brain Aneurysm Foundation Carol W. Harvey Chair of Research, Sharon Epperson Chair of Research, Project Title: A Whole Blood RNA Diagnostic for Unruptured Brain Aneurysm: Risk Assessment Prototype Development and Testing

Financial Interest/Investor/Stock Options/Ownership:
3N Endovascular, Adona Medical, Inc., Basecamp Vascular SAS, Bend IT Technologies, Ltd., BlinkTBI, Inc, Borvo Medical, Inc., CerebrovaKP, Code Zero Medical, Inc., Cognition Medical, Collavidence, Inc., Contego Medical, Inc., CVAID Ltd., E8, Inc., Endostream Medical, Ltd, FreeOx Biotech, SL, Galaxy Therapeutics, Inc., Hyperion Surgical, Inc., Imperative Care, Inc., InspireMD, Ltd., Instylla, Inc., IRRAS, AB, Launch NY, Inc., Neurolutions, Inc., Neurovascular Diagnostics, Inc., NeXtGen Biologics, Peijia Medical, PerFlow Medical, Ltd., Physician X, LLC, Piraeus Medical, Inc., Prometheus Therapeutics, Inc., PUMA Venture Capital Fund I, LP, Q'Apel Medical, Inc., QAS.ai, Inc., Radical Catheter Technologies, Inc., Rist Neurovascular, Inc. (Purchased 2020 by Medtronic), Sense Diagnostics, Inc., Serenity Medical, Inc., Silk Road Medical, Sim & Cure, Spinnaker Medical, Inc., Stent'Up, StimMed, LLC, Synchron, Inc., Tegus Medical, GmbH, T.G. Medical, Inc., Tulavi Therapeutics, Inc., Vastrax, LLC,, Viseon, Inc., Viz.ai, Whisper Medical, Inc., Willow Medtech, Inc.

Consultant/Advisory Board:
Abbott Laboratories, Asahi Intecc Co. Ltd., Canon Medical Systems USA, Inc., Carotix Medical, LLC, CerebrovaKP, Cerenovus, Contego Medical, Inc., Cordis, Endostream Medical, Ltd, FreeOx Biotech, SL, Hyperfine Operations, Inc., Imperative Care , InspireMD, Ltd., IRRAS AB, Medtronic, Minnetronix Neuro, Inc., Peijia Medical, Perflow Medical, Ltd., Piraeus Medical, Inc., Prometheus Therapeutics, Inc., Q'Apel Medical, Inc., Serenity Medical, Inc., Shockwave Medical, Inc., StimMed, LLC, Stryker Neurovascular., Synchron Australia Pty Ltd., Tegus Medical, GmbH, T.G. Medical, Inc., Terumo Neuro (Formerly MicroVention), Vastrax, LLC, Vesalio, Viz.ai, Inc., WL Gore

National PI/Steering Committees:
Cerenovus EXCELLENT and ARISE II Trial; Medtronic SWIFT PRIME, VANTAGE, EMBOLISE and SWIFT DIRECT Trials; Terumo Neuro (formerly MicroVention) FRED Trial & CONFIDENCE Study; MUSC POSITIVE Trial; Penumbra 3D Separator Trial, COMPASS Trial, INVEST Trial, MIVI neuroscience EVAQ Trial; Rapid Medical SUCCESS Trial; InspireMD C-GUARDIANS IDE Pivotal Trial;

Patents:
Patent No. US 11,464,528 B2, Date: October 11, 2022, CLOT RETRIEVAL SYSTEM FOR REMOVING OCCLUSIVE CLOT FROM A BLOOD VESSEL, Applicant and Assignee: Neuravi Limited (Galway), Role: Co-Inventor

**Ciprian N Ionita**
Founder and Chief Scientific Officer at QAS.AI.inc